\documentclass[11pt,draftcls,onecolumn]{IEEEtran}
\usepackage[cmex10]{amsmath}
\usepackage{cite}
\usepackage{epsfig}
\begin{document}
\title{Yield--Optimized Superoscillations}
\author{Eytan~Katzav$^1$ and~Moshe~Schwartz$^2$\\
$^1$Department of Mathematics, King's College Lonodn, Strand, London, WC2R 2LS, UK\\
$^2$Department of Physics, Raymond and Beverly Sackler Faculty  Exact Sciences, Tel Aviv University, Tel Aviv 69978, Israel

\thanks{$^1$e-mail: eytan.katzav@kcl.ac.uk}
\thanks{$^2$e-mail: bricki@netvision.net.il}
}
\maketitle
\begin{abstract}
Superoscillating signals are band--limited signals that oscillate in some region faster their largest Fourier component. While such signals have many scientific and technological applications, their actual use is hampered by the fact that an overwhelming proportion of the energy goes into that part of the signal, which is not superoscillating. In the present article we consider the problem of optimization of such signals. The optimization that we describe here is that of the superoscillation yield, the ratio of the energy in the superoscillations to the total energy of the signal, given the range and frequency of the superoscillations.  The constrained optimization leads to a generalized eigenvalue problem, which is solved numerically. It is noteworthy that it is possible to increase further the superoscillation yield at the cost of slightly deforming the oscillatory part of the signal, while keeping the average frequency. We show, how this can be done gradually, which enables a trade-off between the distortion and the yield. We show how to apply this approach to non-trivial domains, and explain how to generalize this to higher dimensions.
\end{abstract}


%
\IEEEpeerreviewmaketitle


\IEEEPARstart{S}{uper} oscillatory functions provide a stunning refutation of a very widely accepted lore, that band--limited functions cannot oscillate with a frequency larger than its maximal Fourier component. A number of examples have been given in the past for such functions with very interesting applications to Quantum Mechanics \cite{Aharonov88,Aharonov90,Berry94a,Berry94b,Kempf2000,Berry06,Kempf04}, signal processing \cite{Slepian61,Levi65,Slepian78,Kempf02,Kempf06} and to optics \cite{Berry12a}, where superoscillations are intimately related to superresolution \cite{Zheludov08,Zheludov09,Zalevsky11,Zheludov11,Gazit09,Huang07,Wong12}, to superdirectivity or supergain \cite{Wong10,Berry12b} and actually to compression beyond the Fourier limit \cite{Wong11}. Interestingly, it was discovered that in random functions, defined as superpositions of plane waves with random complex amplitudes and directions, considerable regions are naturally superoscillatory \cite{Dennis08,Dennis09}. Various mathematical aspects of the phenomenon have been discussed more recently in \cite{Aharonov11,Aharonov12}. This field is also closely related to several other subjects such as prolate spheroidal wavefunctions \cite{Slepian61,Slepian78} that can be seen as sets of orthogonal superoscillating functions, and to the stability of band--limited interpolation \cite{Ferreira00,Ferreira01} where the lack of higher harmonics challenges interpolating procedures that wish to recover signals containing such higher frequencies.

In a very important sense, though, the idea that a band--limited function cannot oscillate faster than its largest Fourier component is not entirely false. It is well known that the superoscillations exist in limited intervals of time (or regions of space, depending on the actual problem) and that the amplitude of the superoscillations in those regions is extremely small compared to typical values of the amplitude in non-oscillating regions \cite{Aharonov90,Kempf02,Kempf06}. It is generically so small, that any hope of practical application of superoscillating functions depends on tailoring the functions carefully to reduce that effect as much as possible. Two different approaches have been offered over the years to the problem of optimization of superoscillation~\cite{Slepian61,Kempf06} in the absence of constraints, and one~\cite{Kempf04} in the presence of constraints, which will discussed in relation to ours as we proceed.

In contrast to the last three references we prefer to work with periodic functions, which will thus be described by a finite number of Fourier coefficients. Periodic superoscillating functions have been studied in the past, e.g. in \cite{Wong11}, but in a different direction than the work presented here. The choice of periodic functions, we believe, has a number of advantages. First, it is clear, that it is more difficult to achieve superoscillations, with a finite number of degrees of freedom than with an infinite number of degrees of freedom encoded in the Fourier {\bf transform} of a band--limited non-periodic function. Thus, achieving superoscillations with a finite number of degrees of freedom is more challenging, yet enables a clear view of questions concerning the total number of oscillation versus the number of degrees of freedom. Second, we obtain an easy and practical way of constructing optimal superoscillations. 

In the following we discuss first superoscillations in one finite subinterval. This is not essential, and towards the end we show that generalizing this to more than one subinterval is straightforward and requires no conceptual modifications. We will also comment on higher dimensions.

Consider the function 
\begin{equation}
f(t) = \frac{A_0}{\sqrt{2\pi}} + \sum\limits_{m = 1}^N {\frac{A_m}{\sqrt \pi}} \cos (mt)
\label{eq:1}
\end{equation}
Choose an interval $[0,a]$ with $a<\pi$. Impose on the function $f(t)$ $M$ constraints inside the interval, i.e. $f(t_j)=\mu_j$ for $0\le t_j \le a$  and $j=0,\ldots ,M-1$. The constraints result in a set of $M$ linear equations in $N+1$ unknowns of the form, 
\begin{equation}
\sum\limits_{m = 0}^N C_{jm} A_m \equiv {\bf{C}}_j \cdot \bf{A} = \mu_j \, ,
\label{eq:2}
\end{equation}
where 
\begin{equation}
\setlength{\nulldelimiterspace}{0pt}
C_{jm} = \frac{1}{\sqrt \pi} \left\{\begin{IEEEeqnarraybox}[\relax][c]{l's}
 \cos (m{t_j}),& for $m \ne 0$\\
1/\sqrt 2,& for $m=0$%
\end{IEEEeqnarraybox}\right. \, .
\nonumber
\end{equation}
Note that imposing the $M$ constraints described above generically results in $M$ independent linear equations. However, if these equations are not independent, one can eliminate one (or more) constraints such that these eliminated constraints are satisfied automatically when imposing the other constraints. Therefore, without loss of generality, we will assume in the following that we are dealing with $M$ independent constraints, which yields a non-singular matrix $C_{jm}$ (i.e., a matrix of rank $M$).

Therefore, this set of equations has no solution for $M>N+1$, has one solution for $M=N+1$ and a whole space of solutions for $M<N+1$. In particular, we can choose
\begin{equation}
t_j=\frac{a j}{M} \quad \rm{and} \quad \mu_j=(-1)^j \, .
\label{eq:3}
\end{equation}
Provided $M\le N+1$, this choice constraints the function to oscillate within the interval $[-a,a]$ between the values $\pm 1$ with a frequency  
\begin{equation}
\omega  = \frac{\pi M}{a} \, .
\label{eq:4}
\end{equation}
It is thus clear that the frequency of oscillation within the interval $[-a,a]$ can be increased indefinitely just by decreasing its size. Therefore, although to have a solution at all, we need that $M\le N+1$, the ratio between $\omega$ and $N$, the largest frequency appearing in the Fourier series, can be made as large as we want by decreasing $a$. Thus it is not a problem at all to obtain superoscillations. This comes, at a cost, of course. First, we can obtain superoscillations with a prescribed frequency $\omega$ only within an interval $[-a,a]$ with $a\le \frac{\pi N}{\omega }$ and as stated before and will be demonstrated in the following (see Fig.~\ref{fig:1} below) the amplitude in that region is relatively extremely small. 

Next, we would like to optimize our superoscillating function for fixed $a$ and $M<N+1$ but we have to decide first in what sense do we want to optimize it. Ferreira and Kempf \cite{Kempf04,Kempf06}, consider the energy of the signal, $E=\int\limits_{-\infty }^{\infty }{f^2(t)}dt$, and then use the fact that $f$ is band--limited and minimize the energy under the interpolation constraints (Eq.~(\ref{eq:3})). We believe that for many applications, the right quantity to maximize under the constraints is the superoscillation yield,
\begin{equation}
Y(M,a) = \frac{\int\limits_{-a}^a {f^2(t)dt}}{\int\limits_{-\infty }^\infty {f^2(t)dt}} \, ,
\label{eq:5}
\end{equation}
rather than the total energy. (Note that as will become evident in the following the yield is not just a function of $\omega$ but of $M$ and $a$ separately). For the discrete case described in (\ref{eq:1}), we take instead of the energy, which is infinite, the energy per period. Thus the superoscillation yield that we maximize under the constraints is 
\begin{equation}
Y(N,M,a) = \frac{\int\limits_{-a}^a {f^2(t)dt}}{\int\limits_{-\pi}^\pi {f^2(t)dt} } = \frac{\sum\limits_{m,n = 0}^N {\Delta_{mn} A_m A_n}}{\sum\limits_{m = 0}^N A_m^2} \equiv \frac{I}{D} \, ,
\label{eq:6}
\end{equation}
where the entries of the matrix $\Delta$, which correspond to the choice of the interval $[-a,a]$, are given by
\begin{equation}
\setlength{\nulldelimiterspace}{0pt}
\Delta_{mn} = \left\{\begin{IEEEeqnarraybox}[\relax][c]{l's}
{\textstyle{2 \over \pi }}{\textstyle{{m\cos (na)\sin (ma) - n\cos (ma)\sin (na)} \over {m^2-n^2}}}
,& $m \ne n \ne 0$\\
\frac{1}{\pi }\left( a + \frac{\sin (2na)}{2n} \right),& $m=n \ne 0$\\
\frac{\sqrt{2}}{\pi n}\sin (n a),& $m=0$, $n \ne 0$\\
\frac{\sqrt{2}}{\pi m}\sin (m a),& $n=0$, $m \ne 0$\\
\frac{a}{\pi},& $m=n=0$%
\end{IEEEeqnarraybox}\right. \, .
\label{eq:7}
\end{equation}
Note that a general formula for $\Delta_{mn}$ in any domain is given later in Eq.~(\ref{eq:19}).

The set of $M$ vectors $\{{\mathbf{C}_j}\}$ defined in (\ref{eq:2}) spans an $M$ dimensional space. In this space we introduce an orthonormal basis $\{\hat{e}_{N-M+1},\ldots ,{\hat{e}_N}\}$. An orthonormal basis for the full $N+1$ dimensional vector space is then constructed by adding the set $\{\hat{e}_0,\ldots ,\hat{e}_{N-M}\}$, such that $\hat{e}_i\cdot \hat{e}_j=\delta_{ij}$ for all $i,j=0,\ldots ,N$ (up to the orthogonality requirement, the basis vectors can be chosen randomly, of course). We obtain thus the rotated degrees of freedom,
\begin{equation}
B_i = \hat e_i \cdot {\bf{A}} \, ,
\label{eq:8}
\end{equation}
with the obvious advantage that the last $M$ degrees of freedom in $\bf B$ are constrained independently of each other and are equal to linear combinations of the $\mu_j$'s. Thus we denote 
\begin{equation}
B_i = {\tilde \mu _i} \quad {\rm for} \quad i = N - M + 1, \ldots ,N\, .
\label{eq:9}
\end{equation}
The numerator $I$ in (\ref{eq:6}) can be written now in terms of the rotated degrees of freedom $B$ as  $I=\sum\limits_{m.n=0}^{N} \Delta_{mn} A_m A_n=\sum\limits_{m,n=0}^{N}{{{\Delta }_{mn}}^{(R)}} B_m B_n$, where ${{\mathbf{\Delta}}^{(\mathbf{R})}}=\mathbf{R\Delta}{\mathbf{R}}^{-1}$ , $\mathbf{R}$ being the rotation that takes $\mathbf{A}$ into $\mathbf{B}$. Let us describe the matrix ${{\mathbf{\Delta}}^{(\mathbf{R})}}$ by the following block form
\begin{equation}
{{\bf{\Delta }}^{({\bf{R}})}} = \left( \begin{array}{l}
 {{{\bf{\tilde \Delta }}}_{(N + 1 - M) \times (N + 1 - M)}} \\ 
 {{{\bf{\mathord{\buildrel{\lower3pt\hbox{$\scriptscriptstyle\frown$}} 
\over \Gamma } }}}_{M \times (N + 1 - M)}} \\ 
 \end{array} \right.\left. \begin{array}{l}
 {{\bf{\Gamma }}_{(N + 1 - M) \times M}} \\ 
 {{{\bf{\bar \Delta }}}_{M \times M}} \\ 
 \end{array} \right) \, ,
\label{eq:10}
\end{equation}
where $\overset{\lower0.5em\hbox{$\smash{\scriptscriptstyle\frown}$}}\Gamma$ is the transpose of $\Gamma$. The superoscillation yield expressed in terms of the unconstrained $B$'s is 
\begin{equation}
Y = {\textstyle{{\sum\limits_{m,n = 0}^{N - M} {{{\tilde \Delta }_{mn}}{B_m}{B_n} + 2\sum\limits_{m = 0}^{N - M} {\sum\limits_{\scriptstyle n =  \hfill \atop 
  \scriptstyle N + 1 - M \hfill}^N {{\Gamma _{mn}}{{\tilde \mu }_n}{B_m} + \sum\limits_{\scriptstyle m,n =  \hfill \atop 
  \scriptstyle N + 1 - M \hfill}^N {{{\bar \Delta }_{mn}}{{\tilde \mu }_m}{{\tilde \mu }_n}} } } } } \over {\sum\limits_{m = 0}^{N - M} {B_m^2 + \sum\limits_{m = N + 1 - M}^N {\tilde \mu _m^2} } }}} \, .
\label{eq:11}
\end{equation}
Differentiating the yield with respect to $B_m$ and equating to zero yields
\begin{equation}
{\bf{B}} =  - ({\bf{\tilde \Delta}} - Y{\bf{I}})^{-1} \bf{\Gamma \tilde \mu} \, ,
\label{eq:12}
\end{equation}
where $\mathbf{I}$ is the unit matrix and $\mathbf{\tilde{\mu}}$ is the vector of the ${{\tilde{\mu }}_{j}}$'s.  It is thus clear that the components of the vector $\mathbf{B}$ depend on $I$ and on $D$ only through the ratio $Y=\frac{I}{D}$.
Those components are, by Cramer's rule, a ratio of two determinants. The determinant in the denominator is clearly a polynomial of degree $N+1-M$ in $Y$. For each entry of $\mathbf{B}$ the determinant in the numerator is that of the matrix, obtained from $(\mathbf{\tilde{\Delta }}-Y\mathbf{I})$ by replacing one of the columns by the vector $(\mathbf{\Gamma \tilde{\mu}})$. Therefore, $B_m$, the $m$'th component of the vector $\mathbf{B}$, for which the yield is extremal, is given by the explicit expression
\begin{equation}
B_m = \frac{P_m^{N - M}(Y)}{{P^{N + 1 - M}}(Y)} \, ,
\label{eq:13}
\end{equation}
where  $P_{m}^{N-M}(Y)$ is a polynomial of degree $N-M$ in $Y$, and ${P}^{N+1-M}(Y)$ is a polynomial of degree $N+1-M$. Clearly, an equation for the superoscillation yield can be obtained by plugging the right hand side of (\ref{eq:12}) directly into the right hand side of (\ref{eq:10}).  We prefer, though, a different route that yields a more simplified form for the equation. The expression $C={{D}^{2}}\sum\limits_{m=0}^{N}{{{B}_{m}}}\frac{\partial (I/D)}{\partial {{B}_{m}}}$ is identically zero by the extremum condition. This results in the following simplified equation for $Y$, 
\begin{eqnarray}
&& \sum\limits_{m = 0}^{N - M} {\sum\limits_{n = 1}^M {{\Gamma _{mn}}{{\tilde \mu }_n}P_m^{N - M}(Y)} } = \left( {\sum\limits_{n = 1}^M {\tilde \mu _n^2} } \right)Y{P^{N + 1 - M}}(Y)\nonumber\\
&&- \left( {\sum\limits_{m,n = 1}^M {{{\bar \Delta }_{mn}}} {{\tilde \mu }_m}{{\tilde \mu }_n}} \right){P^{N + 1 - M}}(Y) \, ,
\label{eq:14}
\end{eqnarray}
which is a polynomial equation of degree $N+2-M$. We view each of the roots of this equation as a generalized eigenvalue, because to each one corresponds an $(N+1-M)$--dimensional vector. (In contrast to the traditional eigenvalue problem, the eigenvectors are determined by the inhomogeneous linear equation (\ref{eq:12}) and the number of generalized eigenvalues is $N+2-M$, and those determine $N+2-M$ generalized eigenvectors. This means, of course, that the set of generalized eigenvectors is linearly dependent.) We are naturally interested in the largest generalized eigenvalue that corresponds to the maximal superoscillating yield. The generalized eigenvector corresponding to that solution has in the interval $[-a,a]$ the exact superoscillation frequency imposed by the constraints. In Fig.~\ref{fig:1} we present the superoscillating signal for  $N=10$, $M=5$ and $a=1$.
\begin{figure}[!t]
\centering
\centerline{\includegraphics[width=1.5in]{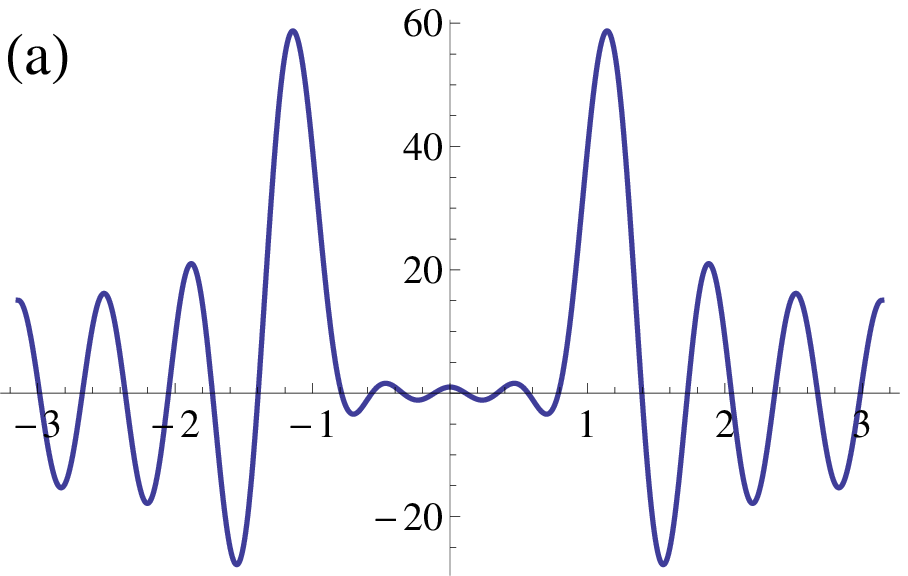} \includegraphics[width=1.5in]{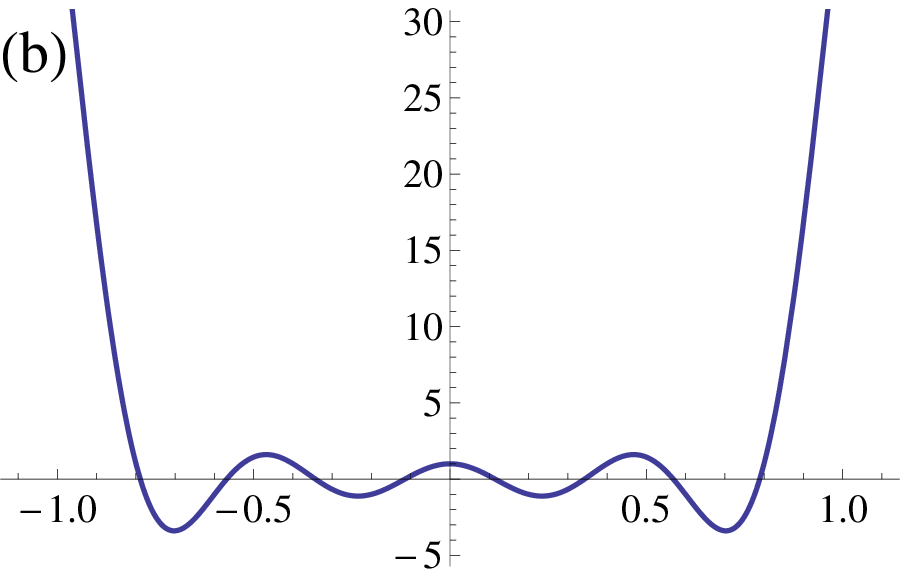}}
\centerline{\includegraphics[width=1.5in]{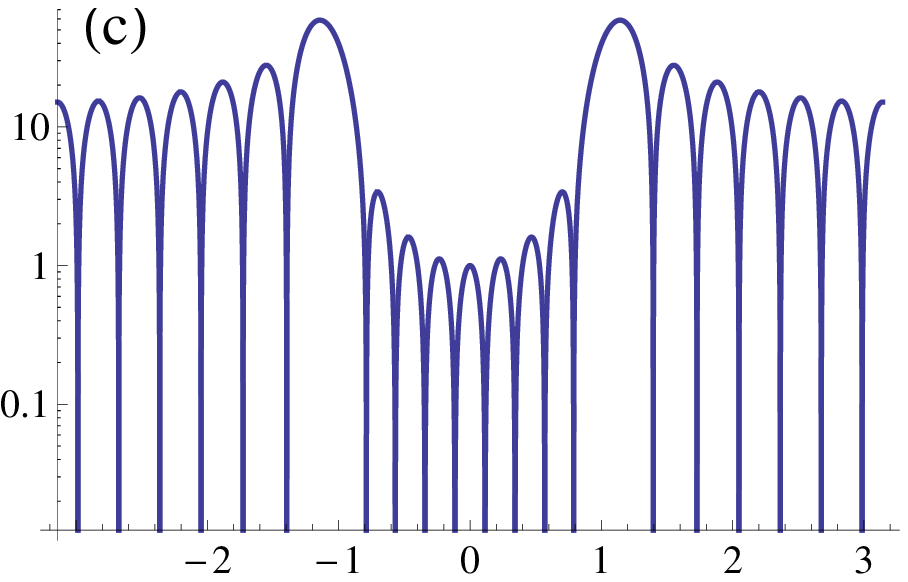}}
\caption{The superoscillating signal for $N=10$, $M=5$ and $a=1$ {\bf (a)} on a linear scale in $(-\pi ,\pi )$, {\bf (b)} on a linear scale focusing on the segment $(-1,1)$, and {\bf (c)} on a semi-logarithmic scale.}
\label{fig:1}
\end{figure}

\begin{figure}[!t]
\centering
\centerline{\includegraphics[width=1.5in]{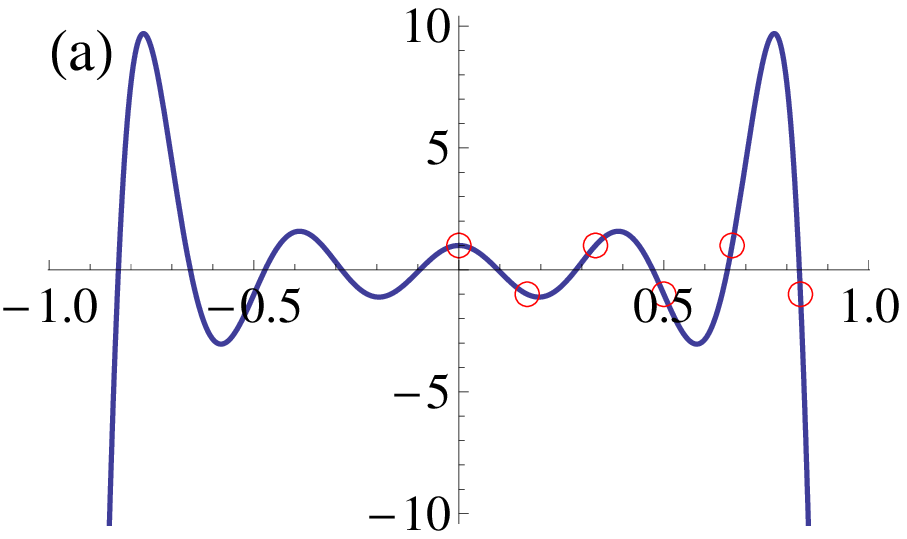} \includegraphics[width=1.5in]{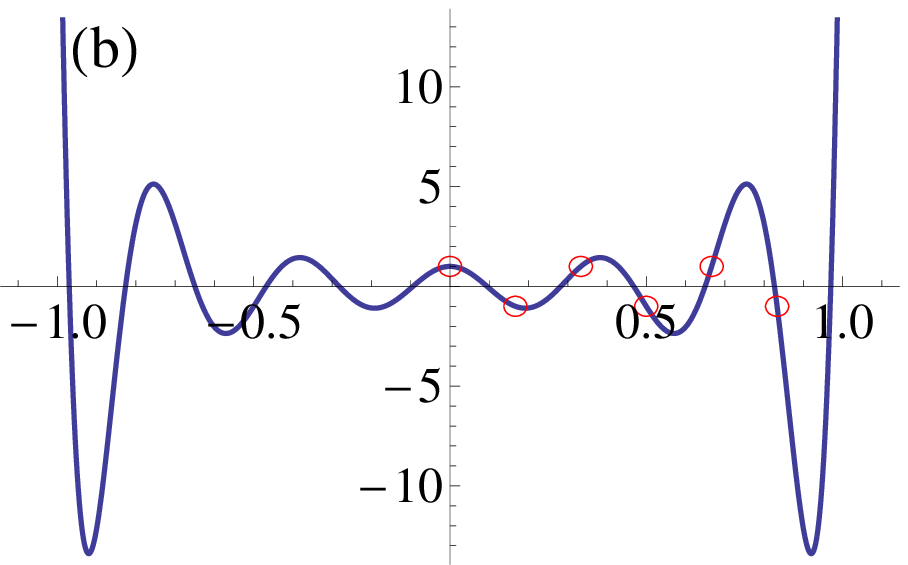}}
\centerline{\includegraphics[width=1.5in]{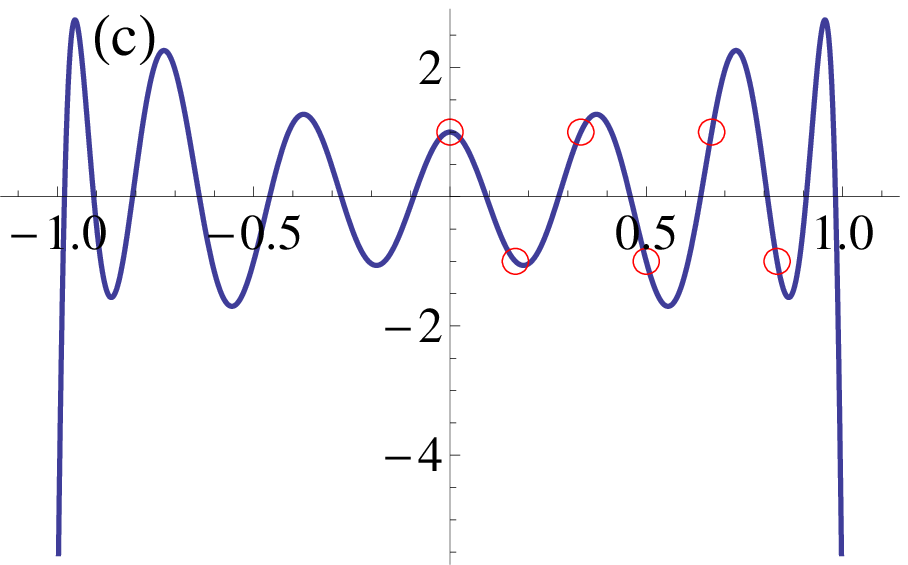} \includegraphics[width=1.5in]{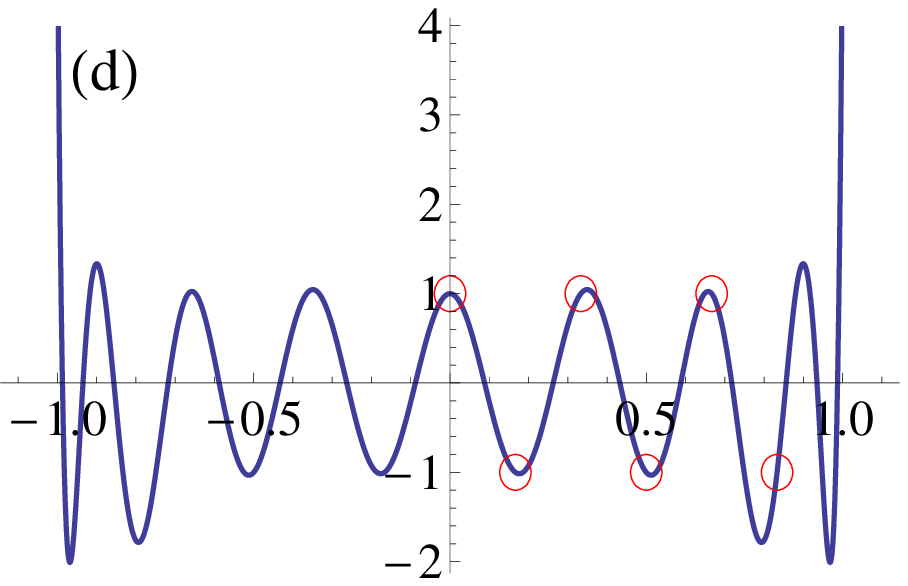}}
\centerline{\includegraphics[width=1.5in]{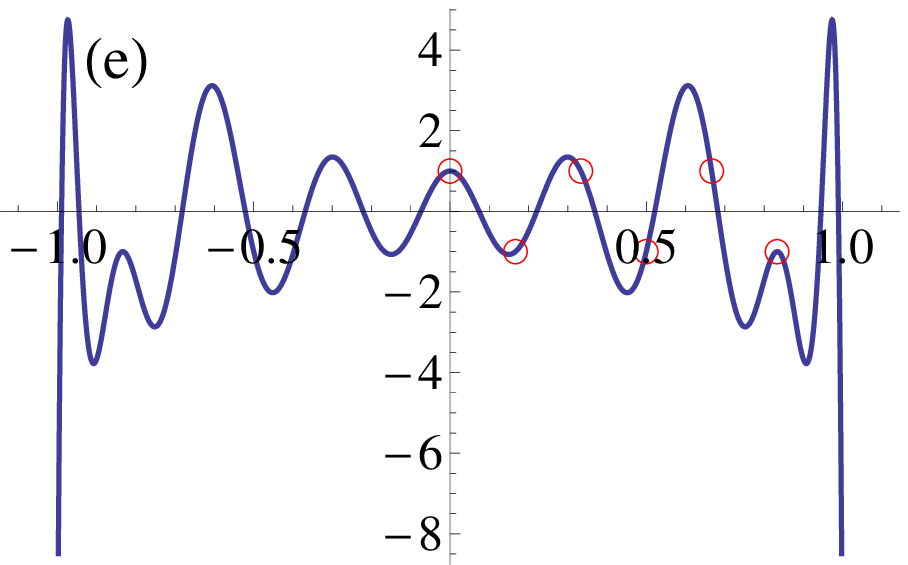} \includegraphics[width=1.5in]{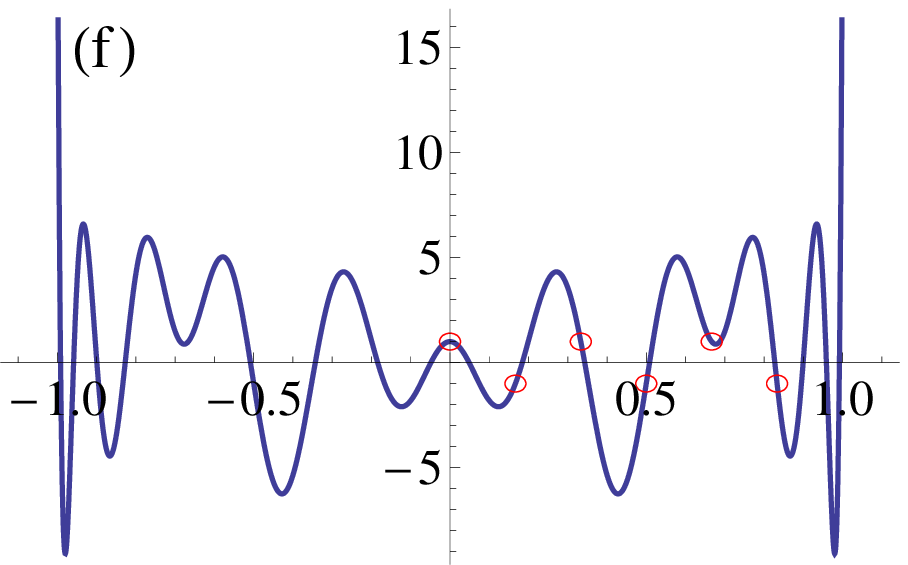}}
\caption{The superoscillating portion of signals corresponding to the various generalized eigenvalues for $N=10$, $M=6$ and $a=2$, with $\lambda_6$ (figure {\bf (a)}) being the maximal eigenvalue, which is the maximal yield and $\lambda_1$ (figure {\bf (f)}) the smallest. The red circles represent the constrained points. The eigenvalues are {\bf (a)} $\lambda_6=0.00233$, {\bf (b)} $\lambda_5=4.847\times 10^{-7}$, {\bf (c)} $\lambda_4=1.379\times 10^{-10}$, {\bf (d)} $\lambda_3=2.559\times 10^{-14}$, {\bf (e)} $\lambda_2=2.176\times 10^{-18}$, {\bf (f)} $\lambda_1=1.189\times 10^{-23}$. As can be seen, every time we switch to a smaller eigenvalue another oscillation inside the superoscillating region appears.}
\label{fig:2}
\end{figure}

A natural question to ask is: can we learn anything from the generalized eigenvectors corresponding to lower generalized eigenvalues? In Fig.~\ref{fig:2} we present the superoscillating portion of the signal corresponding to various generalized eigenvalues for $N=10$, $M=6$ and $a=1$. It is obvious that as we go to lower generalized eigenvalues the number of oscillations in the superoscillating interval keeps increasing, adding more oscillations to those imposed by the constraints. In fact the number of oscillations in the interval $[-a,a]$ grows exactly by one when we go from a generalized eigenvalue $\lambda_i$ to the one immediately below it $\lambda_{i-1} < \lambda_i$.
The number of oscillations inside $[-a,a]$ corresponding to the generalized eigenvalue $\lambda_i$ is thus exactly $N+1-i$ (for $1\le i\le N+2-M$).

\begin{figure}[!t]
\centering
\centerline{\includegraphics[width=3in]{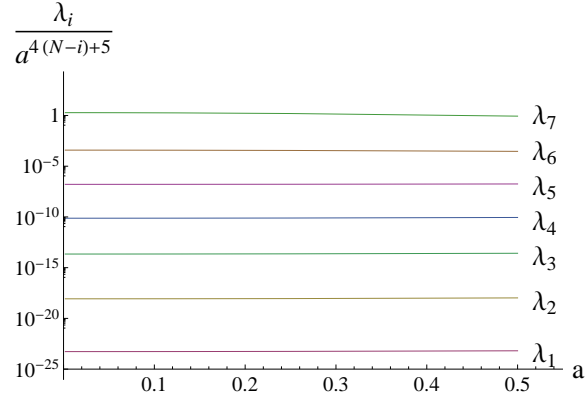}}
\caption{The dependence of the eigenvalue $\lambda_i$ on $a$ for $a<0.5$, fixed $N=10$ and $M=5$. The eigenvalues $\lambda_i$ are divided by $a^{4(N-i)+5}$ to highlight their small-$a$ behaviour.}
\label{fig:3}
\end{figure}

\begin{figure}[!t]
\centering
\centerline{\includegraphics[width=3in]{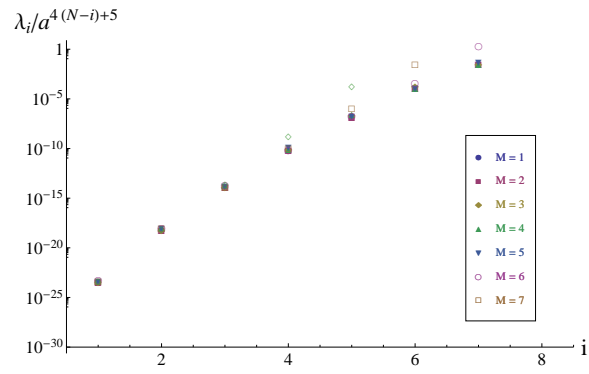}}
\caption{The dependence of the eigenvalue $\lambda_i$ on the index $i$ for fixed $N=10$, $a=1/64$ and various $M$'s.}
\label{fig:4}
\end{figure}

How is $\lambda_i(N,M,a)$ related to the maximal generalized eigenvalue, $Y(N,M,a)$, for the case with the same $N$ and $a$ but with a {\bf constrained} number of oscillations equal to the actual number of oscillation corresponding to $\lambda_i$, $N+1-i$?  The answer seems intuitively clear. Obtaining the same number of oscillations, while imposing less constraints, is expected to give a higher yield. Namely, we expect that for every $M' > M$ and $1\le i\le N+2-M'$ it will hold that $\lambda_i(N,M,a)\ge \lambda_i(N,M',a)$. In particular since $Y(N,M,a)=\lambda_{N+2-M}(N,M,a)$ we also have $\lambda_i(N,M,a)\ge Y(N,N+2-i,a)$. This intuitive feeling, is obviously exact, in the case where the set of $M$ constraints, $S(M)$ obeys $S(M)\subset S(N+2-i)$. To complete the picture for cases where $S(M)\not\subset S(N+2-i)$, we give more details in Figs.~\ref{fig:3} and \ref{fig:4}. In Fig.~\ref{fig:3} we give the different eigenvalues for fixed $N$ and $M$ as a function of $a$. In Fig.~\ref{fig:4} we give the eigenvalue for fixed $a$ and $N$ as a function of $i$ for various values of $M$.  This will give not just an inequality but a full quantitative picture for small (not necessarily very small) $a$. Note that in both figures we divide each eigenvalue $\lambda_i$ by the factor $a^{4(N-i)+5}$ to highlight the small-$a$ behaviour of these eigenvalues. This factor is inspired by the small-$a$ behaviour of the eigenvalues of $\Delta$ (as shown in Refs.~\cite{Ferreira94,Kempf06} for example). The flat nature of the curves in Fig.~\ref{fig:3} verifies this behaviour.

The improvement in the yield for a given number of oscillations in the superoscillation region, obtained by decreasing $i$ comes at a price, though. The seemingly periodic structure near the middle of the superoscillation interval, corresponding to the highest eigenvalue, generically deteriorates as the number of superoscillations increases. This suggests a possible trade-off between the quality of superoscillation and the yield. For that trade-off, we have at our disposal a whole spectrum of signals, with the same number of superoscillations, corresponding to the same number of low Fourier components. On one side of the spectrum we find the signal where all the oscillations are constrained, such that the quality of the superoscillation is good but the yield is relatively low. On the other side of the spectrum we find the function with $M=1$. For that function we only fix the value of $f$ at the origin to be 1. This imposes no oscillation at all on the function in the superoscillating region. Since, optimal $f'$s, which are necessarily symmetric, are not expected to vanish at the origin, this implies that we are not constraining the function at all and the maximization of the yield is equivalent to maximizing it under the requirement that the total "energy" is normalized. This will yield a set of $N+1$ ordinary eigenvalues and eigenvectors. In the corresponding continuum problem the number of eigenvalues is infinite and the eigenvectors are the prolate spheroidal wavefunctions of Slepian and Pollak~\cite{Slepian61}. (Ref.~\cite{Slepian78} mentions a discrete case but it is a totally different discreteness than that we study, i.e., in equation (\ref{eq:1}).) It is interesting to note that the signals obtained by the discrete Ferreira-Kempf procedure~\cite{Kempf04,Kempf06} (who consider a similar question) do not belong to the family of functions described above. Those signals are obtained by minimizing our denominator (defined in (\ref{eq:6})),
\begin{equation}
D = \sum\limits_{m = 0}^N A_m^2  = \sum\limits_{m = 0}^{N - M} B_m^2 + \sum\limits_{m = N + 1 - M}^N {\tilde \mu_m^2} \, ,
\label{eq:15}
\end{equation}
and under the same oscillation constraints we use (i.e., equation (\ref{eq:3})). It is clear thus that those eigenvectors are obtained by simply setting
\begin{equation}
B_m=0 \quad {\rm for} \quad m = 0, \ldots ,N - M \, .
\label{eq:16}
\end{equation}

We now go back and discuss briefly the case of superoscillations in more than one subinterval. Consider the same expansion as in Eq.~(\ref{eq:1}) and let $\cal{D}$ be the domain (which can be in general multiply connected) over which we are interested in having superoscillations. We would like this expansion to be done such that the energy of the signal in $\cal{D}$ will be maximized compared to the total energy of the signal in $(-\pi,\pi)$. This implies maximizing the following superoscillation yield
\begin{equation}
Y\left( N,M,\cal{D}\right) = \frac{\int\limits_{\cal{D}} |f(t)|^2 dt} {\int\limits_{-\pi}^\pi |f(t)|^2 dt} \, ,
\label{eq:17}
\end{equation}
that generalizes Eq.~(\ref{eq:5}). Representing the signal using its Fourier decomposition (\ref{eq:1}) gives
\begin{equation}
Y\left( N,M,\cal{D}\right) = \frac{\int\limits_{\cal{D}} |f(t)|^2 dt} {\int\limits_{-\pi}^\pi |f(t)|^2 dt} = \frac{{\sum\limits_{m,n = 0}^N {{\Delta _{mn}({\cal{D}})}{A_m}{A_n}} }}{{\sum\limits_{m = 0}^N {A_m^2} }} \, ,
\label{eq:18}
\end{equation}
where the matrix $\Delta$ is now $\cal{D}$-dependent, namely
\begin{equation}
\setlength{\nulldelimiterspace}{0pt}
\Delta_{mn} ({\cal{D}})= \left\{\begin{IEEEeqnarraybox}[\relax][c]{l's}
\frac{1}{\pi }\int\limits_{\cal{D}} {\cos \left( {mt} \right)\cos \left( {nt} \right)dt} ,& $m \ne n \ne 0$\\
\frac{1}{\pi }\int\limits_{\cal{D}} {{{\cos}^2}(nt)dt} ,& $m=n \ne 0$\\
\frac{1}{\sqrt 2 \pi}\int\limits_{\cal{D}} {\cos (nt) dt},& $m=0$, $n \ne 0$\\
\frac{1}{\sqrt 2 \pi}\int\limits_{\cal{D}} {\cos (mt) dt},& $n=0$, $m \ne 0$\\
\frac{{Vol\left(\cal{D}\right)}}{2\pi},& $m=n=0$%
\end{IEEEeqnarraybox}\right. \, .
\label{eq:19}
\end{equation}
From this point on, all the approach developed earlier carries exactly the same, with the only difference being that the more general matrix $\Delta_{mn}({\cal{D}})$ is used instead of the one given by Eq.~(\ref{eq:7}).

\begin{figure}[!t]
\centering
\centerline{\includegraphics[width=1.5in]{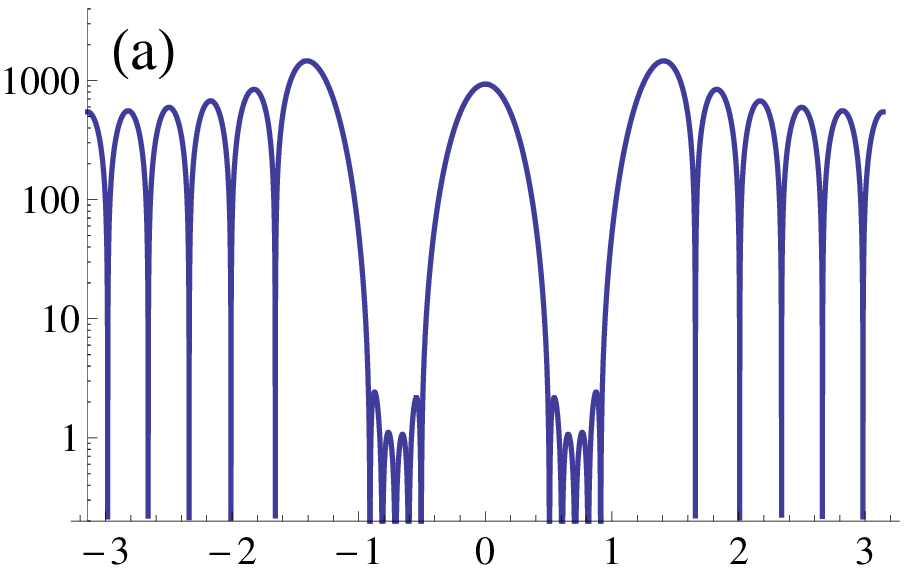} \includegraphics[width=1.5in]{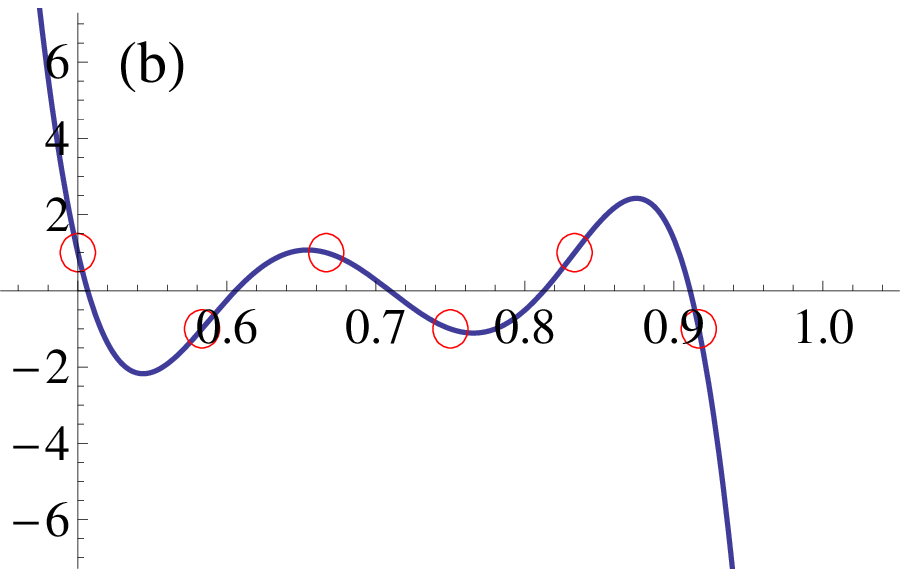}}
\centerline{\includegraphics[width=1.5in]{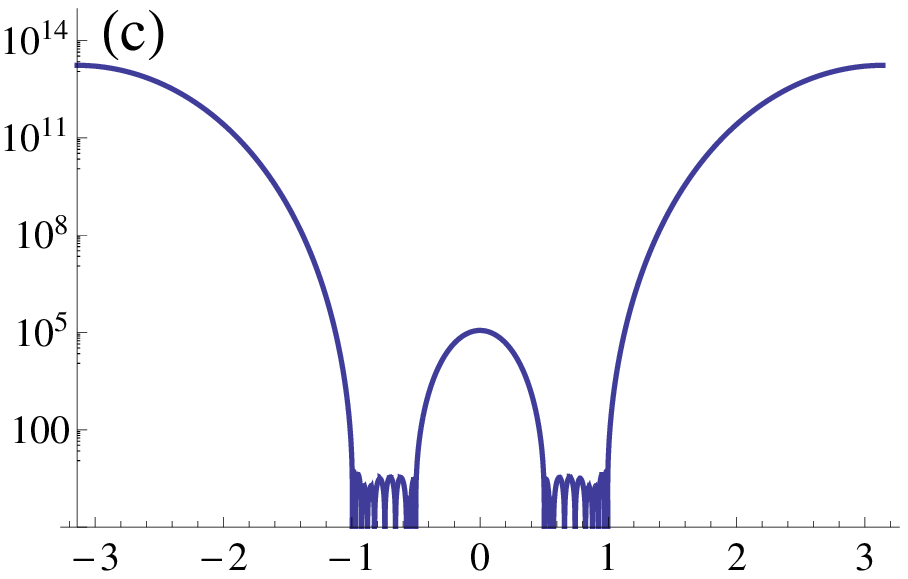} \includegraphics[width=1.5in]{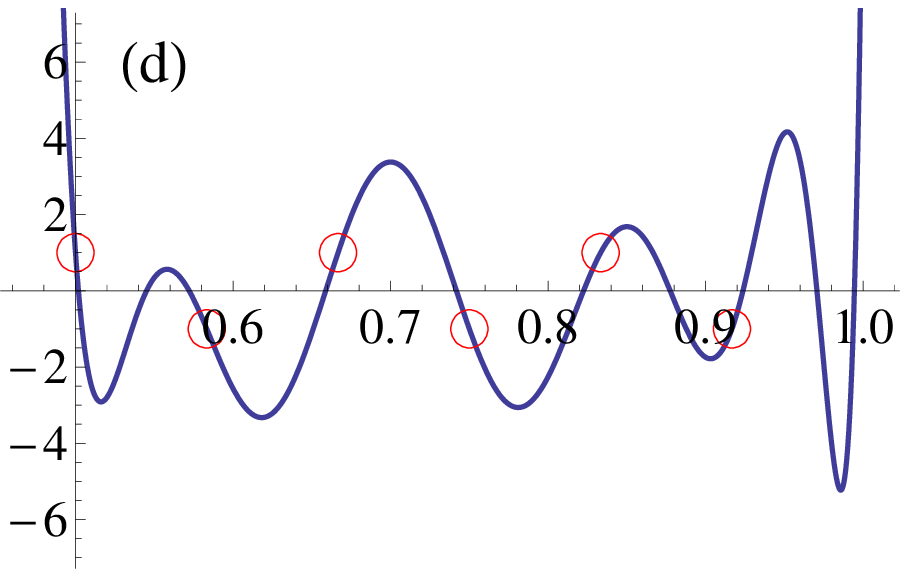}}
\caption{The superoscillating signal for $N=10$, $M=6$ and $(a,b)=(0.5,1)$ corresponding to the largest eigenvalue $\lambda_6=0.000048136$ ({\bf (a)} on a semi-logarithmic scale in $(-\pi ,\pi )$ and {\bf (b)} on a linear scale focusing on the segment $(a,b)$) and the smallest eigenvalue $\lambda_1=2.36786\times 10^{-26}$ ({\bf (c)} on a semi-logarithmic scale in $(-\pi ,\pi )$ and {\bf (d)} on a linear scale focusing on the segment $(a,b)$) . The red circles represent the constrained points, i.e. $f(t_j)=\mu_j$ given by Eq.~(\ref{eq:21}).}
\label{fig:5}
\end{figure}

In order to demonstrate how this works, in the following we will focus on the domain ${\cal{D}}=(-b,-a) \cup (a,b)$ in which case the matrix $\Delta$ becomes
\begin{equation}
\setlength{\nulldelimiterspace}{0pt}
\Delta_{mn} (a,b)= \left\{\begin{IEEEeqnarraybox}[\relax][c]{l's}
{\textstyle{2 \over \pi }}{\textstyle{{ - m \cos(an) \sin (am) + m \cos(bn) \sin(bm) + n \cos (am) \sin (an) - n \cos (bm) \sin (bn)} \over {{m^2} - {n^2}}}},& $m \ne n \ne 0$\\
\frac{1}{\pi }\left[ {\left( {b - a} \right) + \frac{{\sin \left( {2bn} \right) - \sin \left( {2an} \right)}}{{2n}}} \right],& $m=n \ne 0$\\
\frac{\sqrt 2}{\pi} \times \frac{{\sin (bn) - \sin (an)}}{n},& $m=0$, $n \ne 0$\\
\frac{\sqrt 2}{\pi} \times \frac{{\sin (bm) - \sin (am)}}{m},& $n=0$, $m \ne 0$\\
\frac{b-a}{\pi},& $m=n=0$%
\end{IEEEeqnarraybox}\right. \, .
\label{eq:20}
\end{equation}
We impose the following constraints inside $(a,b)$, namely $f(t_j)=\mu_j$ ($j=0,\ldots,M-1$), with
\begin{equation}
t_j=a + \frac{(b - a)j}{M} \quad \rm{and} \quad \mu_j=(-1)^j \, ,
\label{eq:21}
\end{equation}
thus modifying the choice taken in Eq.~(\ref{eq:3}). From this point on, all the procedure presented above is exactly the same as in the case of a single subinterval. In particular Eqs.~(\ref{eq:8})-(\ref{eq:14}) remain valid, once the current choice of $\Delta$ and $t_j$ (that modifies the matrix $C_{jm}$) is used. The result is a set of generalized eigenvalues and eigenvectors, where the largest eigenvalue corresponds to the yield-optimal signal with the given constraints, and every time one takes the next smaller eigenvalue there appears another oscillation inside ${\cal{D}}=(-b,-a) \cup (a,b)$. In Fig.~\ref{fig:5} we present the superoscillating signal for $N=10$, $M=6$ and $(a,b)=(0.5,1)$. We show the signal corresponding to the largest eigenvalue as well as the smallest one. As can be seen visually very clearly, in the case of the smallest eigenvalue all the oscillations are contained in $(a,b)$, thus having even a higher frequency inside this domain, and no oscillations outside. This comes at the expense of of a much smaller energy ratio of course ($\lambda_6=0.000048136$ vs. $\lambda_1=2.36786\times 10^{-26}$). 

Generalizing this discussion to other types of expansions (e.g., an expansion that included the $\sin$ functions) or to higher dimensions (for application see Ref.~\cite{Simons10} for example) is also straightforward, and only requires stating the desired basis for expansion (which fixes the basis functions), the desired domain for superoscillations (which fixes $\Delta$ similar to Eq.~(\ref{eq:19})) and the desired constraints (which fixes the $t_j$'s and $\mu_j$'s). Other than that, the concepts and algorithm itself remain the same.

To conclude, we have shown in this paper how to obtain yield-optimized superoscillating signals that allow a gradual trade-off between superoscillation yield and quality of the signal. In particular we show how the constrained optimization of the energy-ratio can be formulated as a generalized eigenvalue problem. This problem coincides with the standard eigenvalue problem of the operator $\Delta$ in the unconstrained case that give rise to the prolate spheroidal wavefunctions \cite{Slepian61}. The approach presented here allows to shape these wavefunction in ways that may be appealing to applications, such as superresolution~\cite{Zheludov08,Zheludov09,Zalevsky11,Zheludov11,Gazit09,Huang07,Wong12}, compression~\cite{Wong11}, supergain~\cite{Wong10,Berry12b} or band--limited interpolation \cite{Ferreira00,Ferreira01}, where the possibility to improve the shape of such superoscillating signals may turn a beautiful idea to a useful application, which otherwise might remain impractical.

Since our optimization process is based on a specific way of constraining the signal to produce superoscillations, given by equation (\ref{eq:3}), improvements of the yield and/or the superoscillation quality may be expected and will be discussed in future publications, as well as generalizations to other expansion bases and higher dimensions. It is also of interest to obtain rigorous estimates of the optimal yield for superoscillating signals in the presence of many Fourier components ($N >> 1$) and large number of constraints ($M>>1$).

\end{document}